\documentclass[twocolumn,showpacs,preprintnumbers,superscriptaddress]{revtex4}

\usepackage{amsfonts,amssymb,amsmath,amsthm}

\advance\topmargin 2cm

\newtheorem{theorem}{Theorem}

\newtheorem{cor}{Corollary}
\newtheorem{lemma}{Lemma}
\newtheorem{dfn}{Definition}

\newcommand{\CC}{\mathbb{C}}

\newcommand{\calA}{{\cal A }}

\newcommand{\calE}{{\cal E }}

\newcommand{\calM}{{\cal M }}
\newcommand{\calK}{{\cal K }}

\newcommand{\Clif}{{\cal C}}
\newcommand{\Gr}{{\cal G}}
\newcommand{\la}{\langle}
\newcommand{\ra}{\rangle}

\newcommand{\ep}{\epsilon}

\newcommand{\drv}[1]{\frac{\partial}{\partial{#1}}}
\newcommand{\D}[1]{\mathop{\mathrm{D}#1}}
\newcommand{\ad}{{\mathrm{ad}}}

\newcommand{\be}{\begin{equation}}
\newcommand{\ee}{\end{equation}}
\newcommand{\nn}{\nonumber}
\newcommand{\ba}{\begin{array}}
\newcommand{\ea}{\end{array}}

\newcommand{\Pf}[1]{\mathop{\mathrm{Pf}(#1)}}

\newcommand{\aop}{{\hat{a}}}

\newcommand{\cop}{{\hat{c}}}
\newcommand{\nop}{{\hat{N}}}
\newcommand{\Pop}{{\hat{P}}}

\newcommand{\Lop}{{\hat{\Lambda}}}
\newcommand{\Iop}{{\hat{I}}}

\usepackage{graphicx}
\usepackage{dcolumn}
\usepackage{bm}

\DeclareMathOperator*{\tr}{\mathop{\mathrm{Tr}}}

\begin{document}

\title{Lagrangian representation for fermionic linear optics}

\author{Sergey Bravyi}
\affiliation{Institute for Quantum Information,
California Institute of Technology,\\
Pasadena, 91125 CA, USA.}

\date{\today}

\begin{abstract}
Notions of a Gaussian state and a Gaussian linear map are
generalized to the case of anticommuting (Grassmann) variables.
Conditions under which a Gaussian map is trace preserving and (or) completely 
positive are formulated.
For any Gaussian map an explicit formula relating correlation matrices
of input and output states is presented.
This formalism allows to develop the Lagrangian representation for 
fermionic linear optics (FLO). It
covers both unitary operations and the  single-mode projectors
associated with FLO measurements. 
Using the Lagrangian  representation we reduce
a classical simulation of
FLO to a computation of Gaussian integrals over  Grassmann variables.
Explicit formulas describing evolution of a quantum state under FLO
operations are put forward. 
\end{abstract}

\pacs{03.67.-a}

\maketitle

\section{Introduction}
\label{sec:intro}

Photons and electrons are fundamental particles which are widely used
in the modern communication and computing technology.
Since these particles can be used to represent quantum information as well,
they could  be involved in  quantum computing technologies of the future.
It was discovered recently by Knill, Laflamme and
Milburn~\cite{KLM00} that passive linear optical elements together with squeezers,
single photon sources and photo-detectors provide a universal set of operations
for quantum computation. After that discovery Terhal, DiVincenzo~\cite{TV01}, and
Knill~\cite{Knill01} examined a computational capability of
``fermionic linear optics'' (FLO). This term basically refers to
a system consisting of non-interacting electrons in a controllable
external potential and a detector
that measures projectively occupation numbers of single-electron modes. 
It turned out that 
FLO is not a promising platform for building
a universal quantum computer, since it can be efficiently simulated by classical means.
On the other hand, classical simulatibility makes FLO
interesting from the perspective of quantum communication theory, since it might help
to understand better general properties of quantum channels.

As we shall argue, FLO shares many features with a restricted version 
of photon linear optics
where only homodyne measurements are allowed. Using linear optical elements,
squeezers, and homodyne measurements one can generate only Gaussian 
states, and realize only Gaussian maps, see~\cite{GC02}, which also
can be simulated classically. One reason for this similarity is that
non-interacting fermions (bosons) are described in the Lagrangian
representation by Gaussian integrals over anticommuting 
(commuting) variables.  In the present paper
the notions of a Gaussian state and a Gaussian linear map are
generalized to the case of anticommuting (Grassmann) variables.
Conditions under which a Gaussian map is trace preserving and (or) completely 
positive are explicitly formulated.

As was pointed out in~\cite{Knill01}, the weakness of fermions is that all
FLO operations, both the unitaries
and the projectors, 
belong to a closure of a Lie group of relatively small
dimension (growing only quadratically with a number of modes).
We shall call elements of this Lie group
as Gaussian operators. Roughly speaking, any Gaussian operator can
be represented as an exponent of another operator which is quadratic in
the creation/annihilation operators.
 Unfortunately, this definition
can not be applied directly in many important cases, when an operator
does not have a full rank. 
We show how to avoid
problems related with taking a closure and considering
limiting points by defining Gaussian operators in terms of
anticommuting variables and taking advantage of their symmetry properties.
It turns out that the adjoint action of any Gaussian operator 
is a Gaussian linear map.

A set of states that can be
achieved by FLO operations starting from the Fock vacuum is a 
set of fermionic Gaussian states. As it is the case with the bosonic counterpart,
a fermionic Gaussian state can be completely characterized by 
its correlation matrix and thus admits an efficient classical description.          
We shall study an action of Gaussian maps on Gaussian states in details
and describe explicitly how correlation matrices of input and output states
are related to each other. 

A classical simulation algorithm for FLO must be able 
\begin{itemize}
\item
to reproduce statistics of all measurement outcomes, 

\item 
to describe explicitly  a quantum state of the system at each time step. 
\end{itemize}
The first issue has been addressed in the paper~\cite{TV01},
where an explicit formula for the measurement outcomes
probability distribution has been put forward. 
In a more recent paper~\cite{TV04} DiVincenzo and Terhal described a transformation
of FLO states under single-mode measurements in a special case,
when the total number of particles in the system is conserved.
The representation of FLO in terms of Gaussian maps
allows to describe in an explicit form
a transformation of correlation matrices under the whole set of FLO operations.

The rest of the paper is organized as follows. Section~\ref{sec:FLO} contains
a strict definition of FLO.
Section~\ref{sec:Grassmann}
is a brief outline  of  the Grassmann variables formalism. It is
based primarily on the textbook~\cite{DiFrancesco} and the paper~\cite{Soper78}.
In Section~\ref{sec:isom} we use Grassmann variables to introduce
a convenient parametrization of linear operators and formally define
fermionic Gaussian states.
A set of Gaussian operators incorporating Gaussian states, canonical
Bogolyubov transformations, and FLO projectors is defined in Section~\ref{sec:operators}.
Theorem~\ref{theorem:Gopr} therein provides a non-singular characterization
of Gaussian operators. It proves that FLO transformations preserve the set
of Gaussian states. 
A Lagrangian representation for linear maps (i.e. transformations sending 
operators to operators)  is developed in Section~\ref{sec:maps}
where we define a notion of fermionic Gaussian map.  Conditions under which a
map is trace preserving and bistochastic are described. 
Completely positive Gaussian maps are characterized in Section~\ref{sec:CP}.
Finally, in Section~\ref{sec:meas} we describe Gaussian maps implementing
FLO operations and find their action on the level of correlation matrices.

\section{Fermionic linear optics}\label{sec:FLO}

We start from giving a more strict definition of FLO. Instead of talking
about electrons, it is more convenient to define abstract fermionic modes
which can be described by creation and annihilation operators
$\aop_j^\dag$ and  $\aop_j$, $j=1,\ldots, n$, with $n$ being the total number of
modes. They satisfy the Fermi-Dirac
commutation relations, i.e. $\{\aop_j,\aop_k\}=0$ and $\{\aop_j,\aop_k^\dag\}=\delta_{j,k}$.
An occupation number $N_j$ of the $j$-th mode 
becomes an observable $\nop_j=\aop_j^\dag \aop_j$. The vacuum state $|0_F\ra$
satisfies $\aop_j |0_F\ra=0$ for all $j$. The Fock basis is defined by
\[
|N_1,\ldots,N_n\ra = \left( \aop_1^\dag \right)^{N_1}\cdots
 \left( \aop_{\mathstrut n}^{\dag} \right)^{N_n} |0_F\ra.
\]
Here the occupation numbers take values $N_j\in \{0,1\}$ according to the 
Pauli principle. An arbitrary state of the system can be written as a 
superposition of the Fock basis states.

Quantum computation with FLO starts from a Fock basis state
(w.l.o.g. from the vacuum state). It is acted on by
a sequence of ``linear optical'' elements. Unitary elements describe an evolution under
quadratic (in creation/annihilation operators)  Hamiltonians, that is 
\be\label{H_FLO}
H=\sum_{j=1}^n \ep_j \aop_j^\dag \aop_j + H_t + H_s
\ee
where $\ep_j$ represent individual energies of the modes,
\[
H_t = \sum_{1\le j< k \le n} 
t_{jk} \aop_j^\dag \aop_k + \overline{t}_{jk} \aop_k^\dag \aop_j
\]
represents ``tunneling'' processes, and
\[
H_s = \sum_{1\le j< k \le n} 
s_{jk} \aop_j^\dag \aop_k^\dag + \overline{s}_{jk} \aop_k \aop_j.
\]
If the fermions under consideration are electrons, $H_s$ 
may describe an interaction between the system and a bulk $p$-wave superconductor~\cite{Kitaev}.
In this case creation of two extra electrons in the system is balanced by borrowing one
Cooper pair from the superconductor.

Non-unitary ``linear optical'' elements are single-mode measurements. We can
projectively measure an occupation number  $\nop_j$ for any mode $j$.
Depending upon the outcome, the input state is acted on by a projector $\aop_j^\dag \aop_j$ 
or by a projector $\aop_j \aop_j^\dag$. 
In a FLO quantum computation we can alternate between 
measurements and unitary elements, such that   
a choice of the next operation may depend upon the prior  
measurement results.

Instead of $2n$ creation and annihilation operators it will be convenient
to introduce $2n$  Hermitian operators
\be\label{majorana}
\cop_{2j-1} = \aop_j^\dag + \aop_j \quad \mbox{and} \quad
\cop_{2j}=(-i)(\aop_j^\dag - \aop_j).
\ee
They are analogous to coordinate and momentum operators for bosonic 
modes. It follows from the Fermi-Dirac commutation relations that
\be\label{Clifford}
\cop_a \cop_b + \cop_b \cop_a = 2\delta_{ab} \Iop
\ee
for all $1\le a,b \le 2n$. An algebra generated by the operators $\{\cop_a\}$
is called the Clifford algebra and will be denoted $\Clif_{2n}$. An arbitrary
operator $X\in \Clif_{2n}$ can be represented as a polynomial in $\{\cop_a\}$, namely
\[
X= \alpha \Iop + \sum_{p=1}^{2n}\;  \sum_{1\le a_1<\cdots <a_p\le 2n}
\alpha_{a_1,\ldots,a_p} \cop_{a_1}\cdots \cop_{a_p},
\]
where $\alpha=2^{-n}\tr{(X)}$.
We shall call an operator $X\in \Clif_{2n}$ {\it even} ({\it odd}) if it involves only
even (odd) powers of the generators  $\cop$.
Thus the operator algebra describing $n$ fermionic modes coincides with $\Clif_{2n}$.
In particular the Hamiltonian in Eq.~[\ref{H_FLO}] looks as
\be\label{H_2}
H = \frac{i}4 \, \sum_{a,b=1}^{2n} H_{ab} \cop_a \cop_b,
\ee
where $\{H_{ab}\}$ may be an arbitrary antisymmetric real $2n\times 2n$ matrix.
Unitary elements of FLO constitute a group of {\it canonical transformations}
$G_c\subset U(2^n)$, that is $V\in G_c$ iff $V=\exp{(iH)}$ with $H$ from Eq.~[\ref{H_2}].
Adjoint action of a canonical transformation $V\in G_c$ on the operators $\{\cop_a\}$
is just a rotation:
\be\label{rotation}
V \cop_a V^\dag = \sum_{b=1}^{2n} R_{ab} \cop_b, \quad
RR^T =I, \quad \det{(R)}=1.
\ee
By choosing appropriate $H$ one can implement an arbitrary rotation $R\in SO(2n)$.

\section{Anticommuting variables}\label{sec:Grassmann}

It has been known since the late seventies that a powerful technique to study
systems with infinite number of fermionic modes (fermionic quantum fields)
is the functional-integral, or Lagrangian representation (LR). In this representation
many  interesting quantities such as the Green's function or the partition function
are expressed in terms of Feynman integrals over Grassmann valued field.
A concise exposition of this technique can be found in the original
paper by Soper~\cite{Soper78}. Of course  LR can be
applied to a system  with finite number of modes as well. 

In the present paper we slightly customize LR to make it more suitable for
description of FLO. 
First of all, since the unitary elements of FLO
do not preserve the total number of particles, we do not have a preferred vacuum state.
Accordingly, we do not have a preferred normal ordering of fermionic operators.
Secondly, we would like to treat both pure and mixed states on equal footing.
It makes more convenient to develop LR for mixed states and for linear maps
sending mixed states into mixed states, rather than for pure states and
for linear operators sending pure states into pure states.
For the sake of completeness, we start from a brief outline
of the anticommuting variables formalism, see a textbook~\cite{DiFrancesco}
for more details.

Consider $n$-dimensional complex linear space $\CC^{n}$ and denote 
$\theta_1,\ldots,\theta_{n}$ its basis vectors.
Grassmann algebra with complex coefficients $\Gr_{n}$ is 
generated by $\theta_1,\ldots, \theta_{n}$ considered as 
formal variables  subject to multiplication rules
\be\label{Grassman}
\theta_a^2=0 \quad \mbox{and} \quad \theta_a\theta_b + \theta_b\theta_a =0.
\ee
An arbitrary element of $f\in \Gr_{n}$ can be written as a polynomial
of $\theta$'s, namely
\[
f(\theta)=\alpha +\sum_{p=1}^{n}\,
\sum_{1\le a_1<\cdots <a_p\le n}
\alpha_{a_1,\ldots,a_p}\, \theta_{a_1}\cdots \theta_{a_p},
\]
where the coefficients $\alpha_{*}$ are complex numbers.
In other words $\Gr_{n}$ is an antisymmetric tensor algebra over $n$-dimensional
complex space.
A polynomial $f(\theta)$ will be called {\it even} if it involves
only even powers of  $\theta$. Even elements constitute the center
of the Grassmann algebra.

One can formally differentiate functions of Grassmann variables. A partial
derivative over $\theta_a$ is a linear operator 
\[
\drv{\theta_a} \, : \, \Gr_{n} \to \Gr_{n}
\]
which is defined by relations
\[
\drv{\theta_a} 1 =0, \quad \drv{\theta_a}\theta_b =\delta_{ab}
\]
and by the Leibniz's rule
\be\label{Leibniz}
\drv{\theta_a} \left( \theta_b f(\theta)\right) =
\delta_{ab} f(\theta) -\theta_b\drv{\theta_a}f(\theta).
\ee
It follows from this definition that partial derivatives must anticommute:
\be\label{dd=0}
\drv{\theta_a}\drv{\theta_b} + \drv{\theta_b}\drv{\theta_a}=0.
\ee

Since a derivative $\drv{\theta_a} f(\theta)$ does not depend upon variable
$\theta_a$, it is sometimes convenient to think about differentiation as
a linear operator mapping $\Gr_{n}$ into $\Gr_{n-1}$ (strictly speaking, 
after the differentiation one should rename variables $\theta_{a+1},
\ldots,\theta_{n}$ into $\theta_{a},\ldots,\theta_{n-1}$). 
Such an operator is called integration and is denoted as
\[
\int d{\theta_a} \equiv \drv{\theta_a} \, : \, \Gr_{n}\to \Gr_{n-1}.
\]
We will also use a notation 
\[
\int \D{\theta} \equiv \int d\theta_{n} \cdots \int d\theta_2 \int d\theta_1.
\]
The order here is chosen such that $\int\D{\theta}\theta_1\cdots\theta_n=1$.
It follows from Eq.~[\ref{dd=0}] that
\[
\int \D{\theta} \drv{\theta_a} f(\theta) =0,
\]
so by applying the Leibniz's rule Eq.~[\ref{Leibniz}] one can use
anticommuting version of integration by parts.

Under a change of variables
\be\label{Jacobian0}
\eta_a = \sum_{b=1}^n T_{ab} \theta_b, \quad T\in GL(n,\CC)
\ee
operators of differentiation and integration change as follows:
\be\label{Jacobian}
\drv{\eta_a} = \sum_{b=1}^{2n} \left( T^{-1} \right)_{ba} \drv{\theta_b},
\ee
and
\be\label{Jacobian'}
\D{\eta} = \left( \det{T}\right)^{-1} \, \D{\theta}.
\ee

In the rest of this section we consider only even-dimensional Grassmann algebras
$\Gr_{2n}$. Let us consider a quadratic form
\[
\theta^T M \theta \equiv \sum_{a,b=1}^{2n} M_{ab} \theta_a \theta_b \in \Gr_{2n}
\]
defined for any complex antisymmetric $2n\times 2n$ matrix $M$ and
a bilinear form
\[
\theta^T \eta \equiv \sum_{a=1}^{2n} \theta_a \eta_a \in \Gr_{4n}.
\]
Throughout the paper we shall extensively use these two formulas for Gaussian integrals:
\be\label{Gauss1}
\int\, \D{\theta} \exp{\left( \frac{i}2 \theta^T M \theta \right)}
= i^n \Pf{M},
\ee
and 
\begin{eqnarray}
\label{Gauss2}
\int\, \D{\theta} \exp{\left(\eta^T\theta  +  \frac{i}2 \theta^T M \theta \right)} &=&
i^n \Pf{M} \nn \\
&& {} \cdot \exp{\left( -\frac{i}2\, \eta^T M^{-1} \eta \right)}. \nn \\
\end{eqnarray}
In these formulas $\Pf{N}$ is the Pfaffian of a complex antisymmetric matrix $N$ defined as
'antisymmetrized product' $\calA(N_{1,2}N_{3,4}\cdots N_{2n,2n-1})$ that is
\[
\Pf{N}=\frac1{2^n n!} \sum_{\sigma \in S_{2n}} \mbox{sgn}\,{(\sigma)}\,
N_{\sigma_1,\sigma_2} \cdots N_{\sigma_{2n-1}\sigma_{2n}}.
\]
Note that the integral in Eq.~[\ref{Gauss2}] is an element of $\Gr_{2n}$.

\section{Gaussian states}
\label{sec:isom}

To any linear operator $X\in \Clif_{2n}$ one can naturally assign a
polynomial $\omega(X,\theta)\in \Gr_{2n}$ of $2n$ Grassmann variables
defined by
\[
\omega(\cop_p\cop_q\cdots \cop_r,\theta)=\theta_p\theta_q\cdots \theta_r,
\quad \omega(\Iop,\theta)=1.
\]
This definition extends by linearity to an arbitrary $X\in \Clif_{2n}$.
We shall call $\omega(X,\theta)$ a {\it Grassmann representation} of the
operator $X$.
It should be emphasized that $\omega$ is just an isomorphism of linear spaces.
It has nothing to do with multiplication in the algebras $\Clif_{2n}$ and $\Gr_{2n}$.

Consider as an example the projector $\aop_1 \aop_1^\dag$
projecting onto a state ``the first mode is empty''. In terms
of generators $\cop$ one has
\[
\aop_1 \aop_1^\dag = \frac12\left( \Iop + i\cop_1\cop_2\right),
\]
Its Grassmann representation looks as
\be\label{Grassmann_Proj}
\omega(\aop_1 \aop_1^\dag,\theta)=
\frac12\left( \Iop + i\theta_1\theta_2\right) =
\frac12 \exp{ \left(  i \theta_1\theta_2 \right) }.
\ee
Here and below all exponents are defined by their Taylor series.

For any operators $X,Y\in \Clif_{2n}$  one can compute a trace
$\tr(XY)$ using the following simple formula
\be\label{trace}
\tr(X Y) = {(-2)}^n\int D\theta D\mu\,\, e^{\theta^T \mu} 
\omega(X,\theta) \omega(Y,\mu),
\ee
which can be verified by a direct inspection.

Let $V\in G_c$ be a canonical transformation implementing a rotation
$R\in SO(2n)$, see Eq.~[\ref{rotation}], and 
$X\in \Clif_{2n}$ be an
arbitrary operator. One can easily check that
\be\label{ortogonal}
\omega(VXV^\dag,\theta) = \omega(X,\eta),
\quad
\eta_a = \sum_{b=1}^{2n}R_{ab}\theta_b.
\ee
Thus canonical transformations are equivalent to an
orthogonal change of a basis in the space of the Grassmann variables.
We are now ready to define an important class of Gaussian states.
\begin{dfn}\label{def:GS}
A quantum state of $n$ fermionic modes is Gaussian  iff its 
density operator $\rho\in \Clif_{2n}$ has a Gaussian Grassmann
representation, that is
\[
\omega(\rho,\theta)=\frac1{2^n}\exp{\left( 
\frac{i}2 \theta^T M \theta \right)}
\]
for some $2n\times 2n$ real antisymmetric matrix $M$.
The matrix $M$ is called a correlation matrix of $\rho$.
\end{dfn}
Note that $\rho$ itself generally can not be written as an exponent of an
operator, since  it might not have a full rank. By definition
all Gaussian states are described by even polynomial in the Grassmann 
representation, see a remark~\cite{comment}. 
If $\rho\in \Clif_{2n}$ is a Gaussian state, its correlation matrix
can be found from
\[
M_{ab}=\frac{i}2 \tr{\left( \rho [\cop_a,\cop_b] \right)}=
\left\{ \ba{rcl} \tr{\left( \rho\, i\cop_a\cop_b \right)} &\mbox{for}& a\ne b,\\
            0 &\mbox{for}& a=b. \\ \ea \right.
\]
All higher correlators are determined by the Wick formula, namely
\be\label{Wick}
\tr{(\rho\, i^p \cop_{a_1}\cop_{a_2}\cdots \cop_{2p})}=
\Pf{\left. M \right|_{a_1,\ldots, a_{2p}}},
\ee
with  $1\le a_1<\cdots<a_{2p}\le 2n$.  Here $\left. M \right|_{a_1,\ldots, a_{2p}}$ 
is the  $2p\times 2p$ submatrix of $M$ with the indicated rows and columns.
For example,
\[
\tr{(\rho\, i^2 \cop_{1}\cop_{2} \cop_3 \cop_4)}= M_{12} M_{34} - M_{13} M_{24} + 
M_{14} M_{23}.
\]

It is known that any real $2n\times 2n$ antisymmetric matrix $M$ can be
transformed by the adjoint action of $SO(2n)$ into a block-diagonal 
form, with all blocks being $2\times 2$ antisymmetric matrices:
\be\label{M_0}
M = R\, \bigoplus_{j=1}^n
\left( \ba{cc} 0 & \lambda_j \\ -\lambda_j & 0\\ \ea \right) R^T,
\quad R\in SO(2n).
\ee
The absolute values $|\lambda_1|,\ldots,|\lambda_n|$ are referred to as
Williamson eigenvalues of $M$.
It follows from Eq.~[\ref{rotation}] that any Gaussian state $\rho$
can be transformed by a canonical transformation into a product form
\be\label{sep}
\rho(\lambda_1,\ldots,\lambda_n)=
\frac1{2^n} \prod_{j=1}^n (\Iop + i\lambda_j \cop_{2j-1}\cop_{2j}).
\ee
Thus non-negativity of $\rho$ is equivalent to having inequalities 
\[
\lambda_j\in [-1,1], \quad j=1,\ldots, n,
\]
or, in terms of operators, $M^T M \le I$. They are 
necessary and sufficient for a matrix $M$ to be a correlation matrix
of some Gaussian state. The state $\rho$ is pure iff $\lambda_j=\pm 1$,
or $M^T M = I$. Note that the density operator of the Fock vacuum 
is given by Eq.~[\ref{sep}] with all $\lambda_j=1$.

Consider a projective measurement of an occupation number of some 
fermionic mode, say the first one, performed on a Gaussian state
$\rho$ with a correlation matrix $M$.   
Let us   derive in
few lines an explicit expression for the measurement outcomes
probability distribution. The probability of the 
outcome ``the first mode is empty'' is
\[
\tr{(\rho\, \aop_1 \aop_1^\dag)} =
{(-2)}^n 
\int D\theta D\mu\,  e^{\theta^T \mu} \omega(\aop_1\aop_1^\dag,\theta)\omega(\rho,\mu),
\]
see Eq.~[\ref{trace}]. It is convenient to introduce $2n\times 2n$ matrix
$K$ with the only non-zero matrix elements $K_{12}=1$ and $K_{21}=-1$, such that
\[
\omega(\aop_1\aop_1^\dag,\theta)=\frac12 
\exp{\left( \frac{i}2 \theta^T K \theta \right)},
\]
see Eq.~[\ref{Grassmann_Proj}]. Using the 
rules Eq.~[\ref{Gauss1},\ref{Gauss2}] for taking Gaussian integrals,
one gets:
\begin{eqnarray}\nn
\tr{(\rho\, \aop_1 \aop_1^\dag)} &=& 
\frac{{(-1)}^n}{2} \int D\theta\,  e^{\frac{i}2 \theta^T K \theta}
\int D\mu\,  e^{ \theta^T \mu + \frac{i}2 \mu^T M \mu } \nn \\
 &=& 
\frac{(-i)^n}{2} \Pf{M}
\int D\theta\, 
e^{ \frac{i}2 \theta^T (K - M^{-1})\theta  } \nn \\
 &=&
\frac1{2} \Pf{M} \Pf{K - M^{-1}}.\nn \\
\end{eqnarray}
If $M$ is a singular matrix the last expression may be regularized
taking into account that $\Pf{L}^2=\det{(L)}$ for any antisymmetric matrix $L$.
Thus we arrive to
\be\label{prob}
\left[ \tr(\rho\, \aop_1\aop_1^\dag)\right]^2 =
\frac14 \det{\left( M K - I \right)},
\ee
where $I$ is $2n\times 2n$ unital matrix.
A formula similar to Eq.~[\ref{prob}] 
for an overlap between two fermionic Gaussian states
has been known for many years,  see~\cite{Lowdin55}.
In the rest of the paper we will write $X(\theta)$ instead of $\omega(X,\theta)$
for the Grassmann representation of any  operator $X\in \Clif_{2n}$.

\section{Gaussian operators}
\label{sec:operators}

The main goal of this section is to prove that the FLO projectors map
Gaussian states into Gaussian states. This fact has been already proved in
the paper~\cite{Knill01} by demonstrating that
each FLO projector is a limiting point for a converging sequence
of generalized canonical transformations (implementing complex rotations from
the group $SO(2n,\CC)$) and by using continuity arguments. We shall present
an independent proof that relies on a remarkable symmetry of Gaussian states.

\begin{dfn}
\label{def:operators}
${} $

\noindent
(a) An operator $X\in  \Clif_{2n}$ with $\tr(X)\ne 0$ is called Gaussian iff
its Grassmann representation is 
\be\label{Gopr}
X(\theta) = C \exp{\left( \frac{i}2\, \theta^T M \theta \right)}
\ee
for some complex number $C$  and some $2n\times 2n$ complex
antisymmetric matrix $M$. We shall call $M$ a correlation matrix of $X$.

\noindent
(b) An operator $X\in \Clif_{2n}$ with $\tr(X)=0$ is called Gaussian iff
\be\label{Gopr_sing}
X=\lim_{m\to \infty} X_m
\ee
for some converging sequence of Gaussian operators $X_m\in \Clif_{2n}$,
$\tr(X_m)\ne 0$.
\end{dfn}
In other words the set of Gaussian operators is a closure of
a manifold specified by Eq.~[\ref{Gopr}]. 
 For example, Gaussian states
and FLO projectors (see Eq.~[\ref{Grassmann_Proj}]) are Gaussian operators.
As we shall see, canonical transformations are also Gaussian operators,
but some of them have a zero trace and admit only
a singular representation as in Eq.~[\ref{Gopr_sing}] (e.g.
$U=\cop_1\cop_2\in G_c$, $\tr{(U)}=0$).
Roughly speaking, we have a regular representation for density operators and a singular one
for operators describing some transformations. (This is not a matter of concern, 
as these transformations also admit a regular representation in terms of 
Gaussian integrals, see Section~\ref{sec:maps}.) 
Note that all Gaussian operators
are even.

To treat all Gaussian operators on equal footing we shall give an equivalent 
definition which is less explicit but more operational. 
Consider two copies of the original system and an operator
\be\label{Lambda}
\Lop=\sum_{a=1}^{2n} \cop_a\otimes \cop_a,
\ee
which belongs to the tensor product $\Clif_{2n}\otimes \Clif_{2n}$ of two
Clifford algebras~\cite{comment1}. One can easily verify that
$\Lop$ is invariant under any canonical transformation $V\in G_c$ applied
simultaneously to both factors:
\be\label{symmetry}
[\Lop, V\otimes V]=0 \quad \mbox{for any} \quad V\in G_c.
\ee
Besides, a direct inspection shows that $[\Lop,\rho\otimes\rho]=0$
for any 'product' Gaussian state as in Eq.~[\ref{sep}]. 
Since an arbitrary Gaussian state can be converted to a 'product' one 
by a canonical transformation, 
it follows from Eq.~[\ref{symmetry}] that $[\Lop,\rho\otimes\rho]=0$
for any Gaussian state $\rho$.
It turns out that the analogous commutation relation holds for all
Gaussian operators and only for them.

\begin{theorem}
\label{theorem:Gopr}
${}$

\noindent
(1) An operator $X\in \Clif_{2n}$ is Gaussian iff $X$ is even and satisfies
\[
[\Lop, X\otimes X]=0.
\]

\noindent
(2) Any Gaussian operator $X$ has a form
\be\label{Gopr_canon}
X(\theta)=C\left(\prod_{a=1}^{2k} \mu_a\right)
\exp{\left( \frac{i}2 \,
\sum_{a,b=2k+1}^{2n} M_{ab} \mu_a \mu_b 
\right) }
\ee
where $\mu_a = \sum_{b=1}^{2n} T_{ab} \theta_b$ for some invertible
complex matrix  $T$.
\end{theorem}
Let us first discuss several trivial consequences of the theorem.
The part (1) implies that the set of Gaussian operators
is closed under multiplication of operators:
\begin{cor}
If $X,Y \in \Clif_{2n}$ are Gaussian operators then $XY\in \Clif_{2n}$ is 
also a Gaussian operator.
\end{cor}
Since the set of Gaussian operators is closed under taking Hermitian conjugation, 
we infer that 
\begin{cor}
If $X$ is a Gaussian operator and $\rho$ is a Gaussian state
such that $X\rho\ne 0$ then   a state 
\[
\rho' = X\rho X^\dag/ \tr(X\rho X^\dag)
\]
is a Gaussian one. 
\end{cor}
This is quite important result since it tells that an arbitrary
sequence of FLO operations can only produce a Gaussian state,
provided that the initial state was a Gaussian one.
To keep track of the state's evolution one just needs to 
compute a correlation matrix of $\rho'$ given correlation
matrices of $\rho$ and $X$. We shall postpone a solution of this problem
until Section~\ref{sec:maps}.

In the rest of this section we prove Theorem~\ref{theorem:Gopr}.
For convenience the proof is split into two lemmas.

\begin{lemma}\label{lemma:Gopr1}
For any Gaussian operator $X\in \Clif_{2n}$ one has
\be\label{Gopr1}
[\Lop,X\otimes X]=0.
\ee
\end{lemma}
\begin{proof}[Proof:]
Applying the isomorphism of Section~\ref{sec:isom} separately to
both subsystems 
we get 
$\Clif_{2n}\otimes\Clif_{2n}\cong \Gr_{2n}\otimes \Gr_{2n}$,
see~\cite{comment1}. 
Our first goal is to describe the adjoint action 
\[
\Lambda_{\ad}\, : \, O \to [\Lop, O],
\]
which maps $\Clif_{2n}\otimes\Clif_{2n}$ into itself,
in terms of Grassmann variables. Introduce a differential operator
\[
\Delta_a = 2\left( 
\theta_a\otimes \drv{\theta_a} + \drv{\theta_a}\otimes \theta_a\right).
\]
We claim that 
\be\label{Gopr_aux1}
[\cop_a\otimes \cop_a,Y\otimes Z](\theta)=\Delta_a \cdot Y(\theta)\otimes Z(\theta)
\ee
for any operators $Y,Z\in \Clif_{2n}$ having the same parity
(i.e. for both $Y$ and $Z$ being odd or both being even).
Without loss of generality, both $Y$ and $Z$ are monomials in $\cop$'s.
In this case each of them either commutes or anticommutes with $\cop_a$.
Consider two cases. (a) Both $Y$ and $Z$ contain $\cop_a$,
or both $Y$ and $Z$ do not contain
$\cop_a$.  Then the commutator $[\cop_a\otimes \cop_a,Y\otimes Z]$ is zero
since both factors yield the same sign (recall that $Y$ and $Z$ have the
same parity). The righthand side of Eq.~[\ref{Gopr_aux1}] is also zero,
since either $\theta_a$ or $\drv{\theta_a}$ annihilates both $Y$ and $Z$.
(b) $Y$ contains $\cop_a$ while $Z$ does not contain $\cop_a$ (or vice verse).
In this case $\cop_a\otimes\cop_a$ anticommutes with $Y\otimes Z$.
Let us write $Y=\cop_a \tilde{Y}$, where $\tilde{Y}$ is a monomial which does not
contain $\cop_a$. We have:
\[
[\cop_a\otimes \cop_a,Y\otimes Z]=2(\cop_a\otimes \cop_a)(Y\otimes Z) = 2\tilde{Y}\otimes
(\cop_a Z).
\]
On the other hand,
\[
\theta_a\otimes \drv{\theta_a} \cdot Y \otimes Z =0,
\]
and
\[
\drv{\theta_a}\otimes \theta_a \cdot Y\otimes Z =
\tilde{Y}\otimes \theta_a Z.
\]
Substituting the last three formulas into Eq.~[\ref{Gopr_aux1}] we again get an equality.
Summarizing, the adjoint action of $\Lop$ can be described by
a differential operator 
\be\label{Adjoint}
\Lambda_{\ad} = 2\sum_{a=1}^{2n}
\left( 
\theta_a\otimes \drv{\theta_a} + \drv{\theta_a}\otimes \theta_a
\right).
\ee

Now we can easily conclude the proof. If $X$ is a Gaussian operator with 
$\tr(X)\ne 0$, we can write
\[
X(\theta)=\exp{\left( \frac{i}2\, \theta^T M \theta \right) },
\]
(the constant factor $C$ is omitted) and thus
\[
\drv{\theta_a}X  = i\left( \sum_{b=1}^{2n} M_{ab} \theta_b \right) X.
\]
Applying the differential representation of Eq.~[\ref{Adjoint}] we get
\begin{eqnarray}\nn
\Lambda_{\ad}\cdot X\otimes X &=&
2i \sum_{a,b=1}^{2n} (M_{ab} + M_{ba})\, \theta_a\otimes \theta_b \nn \\
&& {} \cdot X\otimes X =0. \nn \\
\end{eqnarray}
Thus $[\Lambda,X\otimes X]=0$. If $X$ is a limiting point for a sequence
of Gaussian operators $\{X_m\}$, $\tr(X_m)\ne 0$, continuity arguments show that
$[\Lambda,X\otimes X]=0$ as well.
\end{proof}

In the next lemma we prove the reverse statement and as a byproduct 
derive an explicit expression Eq.~[\ref{Gopr_canon}] for traceless Gaussian operators.
\begin{lemma}\label{lemma:Gopr2}
Suppose $X\in \Clif_{2n}$ is an even operator such that
$[\Lop,X\otimes X]=0$. Then $X$ is a Gaussian operator.
\end{lemma}

\begin{proof}[Proof:]
Let us first prove the lemma for the case $\tr(X)\ne 0$.
Making use of Eq.~[\ref{Adjoint}] we can rewrite the equality
$[\Lop,X\otimes X]=0$ as
\be\label{Gopr_aux2}
\sum_{a=1}^{2n} 
\left( 
\theta_a\otimes \drv{\theta_a} + \drv{\theta_a}\otimes \theta_a
\right)
X\otimes X =0.
\ee
Denote $C=2^{-n}\tr{(X)}\equiv X(0)$ and represent $X(\theta)$ as
\[
X(\theta)= C\cdot 1 + \frac{iC}2\sum_{a,b=1}^{2n} M_{ab}\,\theta_a \theta_b + 
\mbox{higher order terms}.
\]
Applying a differential operator $1\otimes \drv{\theta_b}$ to both sides
of Eq.~[\ref{Gopr_aux2}] and making use of the Leibniz's rule Eq.~[\ref{Leibniz}]
one gets:
\begin{eqnarray}
\sum_{a=1}^{2n} \left(
\theta_a X \otimes 
\frac{\partial^2}{\partial \theta_b \partial \theta_a} X -
\drv{\theta_a} X \otimes \theta_a\drv{\theta_b} X \right) && \\ \nn
{} + \drv{\theta_b} X \otimes X &=& 0. \nn
\end{eqnarray}
Now let us put $\theta\equiv 0$ in the second factor 
(e.g. pick up a coefficient before 1). Taking into account the
decomposition for $X(\theta)$ given above one easily gets:
\be\label{lin_dif_eq}
\drv{\theta_b} X = i\sum_{a=1}^{2n} M_{ba} \theta_a X.
\ee
This differential equation can be easily solved by making a
canonical transformation which brings $M$ to the block-diagonal
form, see Eq.~[\ref{M_0}], and using formulas Eq.~[\ref{Jacobian}].
Taking into account ``initial conditions'' $X(0)=C$ one gets
\[
X(\theta)=C \exp{\left( \frac{i}2\, \theta^T M \theta \right) },
\]
that is $X$ is a Gaussian operator.

Now let us prove the lemma for the general case.
Consider a linear subspace $\calM_1\subseteq \Gr_{2n}$ spanned by
linear functions of Grassmann variables:
\[
\calM_1=\{f\in \Gr_{2n}\; : \;
f(\theta) = \sum_{a=1}^{2n} \alpha_a \theta_a\}.
\]
Here the coefficients $\alpha_a$ are complex numbers.
Denote $\calK\subseteq \calM_1$ a subspace spanned by linear functions
which annihilate $X$, i.e.
\[
\calK=\left\{ f\in \calM_1 \; : \; f(\theta)X(\theta)=0\right\}.
\]
(Since $X$ is an even operator, any linear function $f\in \calM_1$ commutes with $X$,
so the left annihilation is equivalent to the right annihilation.)
Let us perform a linear change of variables 
\be\label{mu}
\mu_a=\sum_{b=1}^{2n} T_{ab} \theta_b, 
\ee
with $T$ being an invertible complex matrix chosen
such  that the first $k$ variables $\mu$ span the subspace $\calK$, i.e.
\[
\calK=\mbox{linear span}\, [ \mu_1,\ldots,\mu_k].
\]
From equalities $\mu_j X=0$, $j\in [1,k]$ it follows that 
\be\label{mu_factor}
X(\theta(\mu))=\left(\prod_{a=1}^k \mu_a \right) \tilde{X}(\mu),
\ee
where $\tilde{X}(\mu)$ depends only upon $\mu_{k+1},\ldots,\mu_{2n}$.
From Eq.~[\ref{mu}] we infer that
\[
\theta_a = \sum_{b=1}^{2n} (T^{-1})_{ab} \mu_b, \quad
\drv{\theta_a} = \sum_{b=1}^{2n} T_{ba} \drv{\mu_b},
\]
see Eq.~[\ref{Jacobian}].
It implies that the differential operator $\Lambda_{\ad}$ is invariant under
the change of variables:
\[
\Lambda_{\ad}=2\sum_{a=1}^{2n} 
\left( \mu_a \otimes \drv{\mu_a} + \drv{\mu_a}\otimes \mu_a 
\right).
\]
Thus the function $\tilde{X}(\mu)$ satisfies an equation
\be\label{tildeX}
 \sum_{a=k+1}^{2n}
\left( 
\mu_a\otimes \drv{\mu_a} + \drv{\mu_a}\otimes \mu_a
\right)\, \tilde{X}\otimes \tilde{X}=0.
\ee
Note that polynomials $\mu_{a}\tilde{X}$, $a\in[k+1,2n]$ are
linearly independent, since otherwise we could extend the subspace $\calK$.
Then it follows from Eq.~[\ref{tildeX}] that the derivatives 
$\drv{\mu_a} \tilde{X}$ must be some linear combinations of 
polynomials $\mu_a \tilde{X}$, $a\in [k+1,2n]$. 
That is $\tilde{X}(\mu)$ obeys a differential equation similar to Eq.~[\ref{lin_dif_eq}]
and we conclude that
\[
\tilde{X}(\mu) = C \exp{\left( \frac{i}2 \,
\sum_{a,b=k+1}^{2n} \mu_a M_{ab} \mu_b 
\right)}
\]
for some complex $(2n-k)\times (2n-k)$ matrix $M$ and some complex number $C$.
Since $X$ was supposed to be an even operator we infer that $k$ must be an 
even integer. Combining the formula above for $\tilde{X}(\mu)$ and Eq.~[\ref{mu_factor}]
we arrive to the desired representation Eq.~[\ref{Gopr_canon}]
for the operator $X$.

If $k=0$ the lemma has been already proved. If $k>0$, $X$ is a limiting point
for a sequence of regular Gaussian operators
as in Eq.~[\ref{Gopr}] since
\[
\prod_{a=1}^k \mu_a =\lim_{t\to \infty} t^{-\frac{k}2} \exp{\left( t\mu_1\mu_2 +
\cdots + t\mu_{k-1}\mu_k \right)}.
\]
We have proved that $X$ is a Gaussian operator.
\end{proof}

\section{Gaussian linear maps}
\label{sec:maps}

By a linear map we shall mean a linear transformation sending operators
to operators.
A transformation of a quantum state under any 
physical operation (or under a conditioned physical operation such as
a projector) can be described in terms of completely positive (CP) linear
map.  We have shown in the previous section (see Corollary~2
next to Theorem~1) that the adjoint action of any Gaussian operator
is a CP map that has the following nice property: {\it Gaussian operators 
are transformed into Gaussian operators.} 
The main goal of this section is to introduce more general
class of CP maps sharing this property and to describe explicitly
their action on the level of correlation matrices.

\begin{dfn}
\label{def:Gauss_int}
A linear map $\calE \, : \, \Clif_{2n} \to \Clif_{2n}$ is 
Gaussian iff it admits an integral  representation
\be\label{Gauss_int}
\calE(X)(\theta) = C \int \exp{\left[ S(\theta,\eta) + i\eta^T \mu \right]}
X(\mu) \D{\eta}\D{\mu},
\ee
where
\be\label{Action}
S(\theta,\eta)= \frac{i}2 (\theta^T,\eta^T)
\left( \ba{cc} A & B \\ -B^T & D \\ \ea \right)
\left( \ba{c} \theta \\ \eta \\ \ea \right)
\ee
for some complex  $2n\times 2n$ matrices $A$, $B$, $D$, and some
complex number $C$.
\end{dfn}
The function $S(\theta,\eta)$ will be called {\it an action} for the map $\calE$.
Since the matrices $A$ and $D$ enter into Eq.~[\ref{Action}]
as $\theta^T A \theta$ and $\eta^T D \eta$, we can assume them to be
antisymmetric. Let us emphasize that no restrictions are put here on the input
operator $X$; it may be non-Gaussian and need not to have a definite
parity. The double integration in Eq.~[\ref{Gauss_int}] may be thought of as
a trace over one copy of the system, see Eq.~[\ref{trace}].

As a simple example  let us consider the identical map
$\calE_I(X)\equiv X$. One can easily verify that its integral representation
is as follows:
\[
\calE_I(X)(\theta)=X(\theta) = 
\int \prod_{a=1}^{2n} (\theta_a - \mu_a) X(\mu) \D{\mu}.
\]
Taking into account that 
\be\label{delta-kernel}
\prod_{a=1}^{2n} (\theta_a - \mu_a) = 
\int \exp{\left[ i(\theta - \mu)^T \eta \right]} \D{\eta},
\ee
we finally get:
\[
X(\theta) = 
\int 
\exp{ \left[ i\theta^T \eta + i\eta^T \mu \right]} X(\mu) \D{\eta} \D{\mu}.
\]
Thus  $\calE_I$ is a Gaussian map with an action 
$S_I(\theta,\eta)=i\theta^T \eta$,
i.e.  $A=D=0$ and $B=I$.

A canonical transformation implementing a rotation $R\in SO(2n)$
of the generators $\cop$, see Eq.~[\ref{rotation}],
is described by an action $S_R(\theta,\eta)=i\theta^T R^T \eta$, that is
$A=D=0$ and $B=R^T$. Indeed, according to the previous paragraph,
it specifies a linear map $\calE_R$ such that
$\calE_R(X)(\theta)= \calE_I(X)(R\theta)=X(R\theta)$.
This orthogonal change of variables is exactly the effect of
canonical transformations in the Grassmann representation, see Eq.~[\ref{ortogonal}]. 

Now we shall establish several useful properties of Gaussian maps.
Recall that an operator $X\in \Clif_{2n}$ is called even (odd) if 
it is a linear combination of only even (odd) monomials in the generators $\cop$.
\begin{dfn}
A linear map $\calE\, : \, \Clif_{2n}\to \Clif_{2n}$  is
called parity preserving iff it maps even (odd) operators into
even (odd) operators.
\end{dfn}
\begin{lemma}\label{lemma:parity}
Any Gaussian map is  parity preserving.
\end{lemma}
\begin{proof}[Proof:]
For any operator $X\in \Clif_{2n}$ define an operator $\overline{X}\in \Clif_{2n}$
such that $\overline{X}(\theta)=X(-\theta)$. Obviously, $X$ is an even (odd) operator
iff $\overline{X}=X$ ($\overline{X}=-X$). Let us make a change of variables
$\mu\to -\mu$ and $\eta\to -\eta$ in the integral Eq.~[\ref{Gauss_int}].
Taking into account that $S(\theta,-\eta)=S(-\theta,\eta)$ and 
Eq.~[\ref{Jacobian'}] saying that $\D(-\mu)=\D(\mu)$, $\D(-\eta)=\D(\eta)$
we arrive to
\be\label{pp}
\overline{\calE(X)}=\calE(\overline{X}).
\ee
Thus if $\overline{X}=\pm X$ then $\overline{\calE(X)}=\pm \calE(X)$. The lemma is proved.
\end{proof}

\begin{lemma}\label{lemma:Gmap}
Gaussian maps transform Gaussian operators into Gaussian operators.
\end{lemma}
\begin{proof}[Proof:]
Instead of proving this statement by direct application of 
Eqs.~[\ref{Gauss1},\ref{Gauss2}] for
Gaussian integrals (which is not so simple due to possible singularities), 
we will make use of Theorem~\ref{theorem:Gopr} (part~1) and the differential 
representation Eq.~[\ref{Adjoint}] for the adjoint action of 
the operator $\Lambda$. Since Gaussian maps are parity preserving 
and Gaussian operators are even, one suffices to prove that
\be\label{to_be_tested}
\Lambda_{\ad}\cdot  \calE(X)\otimes \calE(X) =0
\ee
for any Gaussian operator $X$ and any Gaussian map $\calE$.
The proof involves three main ingredients: (i) an identity
\be\label{rule1}
\ba{l}
\sum_{a=1}^{2n}
\left( 
\theta_a\otimes \drv{\theta_a} + \drv{\theta_a}\otimes \theta_a
\right)\cdot 
\exp{\left( i\theta^T B \eta \right) }^{\otimes 2} \\
{} = - \sum_{a=1}^{2n}
\left( 
\eta_a\otimes \drv{\eta_a} + \drv{\eta_a}\otimes \eta_a
\right) \cdot
\exp{\left( i\theta^T B \eta \right) }^{\otimes 2} \\
\ea
\ee
which can be easily verified;  (ii) the Leibniz's rule
\be\label{Leib1}
\drv{\theta_a}(fg)=
\left(\frac{\partial f}{\partial \theta_a}\right) g +
f\left(\frac{\partial g}{\partial \theta_a}\right),
\ee
which is valid for even polynomial $f$ and
arbitrary  $g$; (iii)
the integration by parts formula, i.e.
\be\label{rule2}
\int \left( \frac{\partial f}{\partial \theta_a}\right) g \D{\theta} =
\pm \int f \left( \frac{\partial g}{\partial \theta_a}\right) \D{\theta}.
\ee
This formula is valid if one of the functions $f$ and $g$ is even.
Here the sign '+' stands for even $g$ and the sign '-' stands for even $f$.
(If both functions are even then the integrals on the right and on the left
are both equal zero, so one can choose an arbitrary sign.) 
Introduce an auxiliary polynomial
\[
f(\eta)=
\int \exp{\left[  i\eta^T \mu \right]}
X(\mu) \D{\mu}.
\]
By the same arguments as in the proof of Lemma~\ref{lemma:parity} one can show
that $f$ is even. Let us apply $\Lambda_{\ad}$ to a polynomial
\[
P_X\equiv
\left(
\int
\exp{[S(\theta,\eta)]}f(\eta) \D{\eta}
\right)^{\otimes 2} \in \Gr_{2n}\otimes \Gr_{2n}.
\]
We can move $\Lambda_{\ad}$ from the left to the right applying subsequently
(a) The Leibnitz's rule Eq.~[\ref{Leib1}];
(b) Lemma~\ref{lemma:Gopr1}; (c) The identity Eq.~[\ref{rule1}];
(d) The integration by parts formula Eq.~[\ref{rule2}]; 
(f) The Leibnitz's rule; (g) Lemma~\ref{lemma:Gopr1}.
After these steps one gets
\[
\Lambda_{\ad} P_X=
-\int \exp{[S(\theta,\eta)]}^{\otimes 2}
\Lambda_{\ad} f(\eta)^{\otimes 2}
\D{\eta}\otimes \D{\eta}.
\]
Let us keep moving $\Lambda_{\ad}$ to the right by applying subsequently
the identity Eq.~[\ref{rule1}] and the integration by parts formula
Eq.~[\ref{rule2}]. After these moves one gets:
\[
\Lambda_{\ad} f(\eta)^{\otimes 2} =
-\int
\exp{[i\eta^T\mu]}^{\otimes 2}
\Lambda_{\ad}
X(\mu)^{\otimes 2}
\D{\mu}\otimes \D{\mu}.
\]
We arrive to
\[
\Lambda_{\ad}\cdot  \calE(X)\otimes \calE(X)
= \calE \otimes \calE \left( \Lambda_{\ad}\cdot  X\otimes X \right)=0.
\]
The lemma is proved.
\end{proof}

Consider a Gaussian operator $X$ which can be described by a correlation matrix $M$
as in Eq.~[\ref{Gopr}] and a Gaussian map $\calE$ as in Eqs.~[\ref{Gauss_int},\ref{Action}].
Lemma~\ref{lemma:Gmap}  implies that $\calE(X)$ is a Gaussian operator. Applying
the Gaussian integration rule Eq.~[\ref{Gauss2}] one can show that $\calE(X)$
has a correlation matrix
\begin{eqnarray}\label{E(M)}
\calE(M) &=& B \left( M^{-1} + D \right)^{-1} B^T + A \\ \nn
&=& B \left( I + MD \right)^{-1} M B^T +A, \nn
\end{eqnarray}
while a pre-exponential factor of the operator $\calE(X)$ can be found from
an  identity
\be\label{tr(E(M))}
\tr{ \left(\calE(X)\right)} = C (-1)^n \Pf{M} \Pf{M^{-1}+D} \tr{(X)}.
\ee
The value of $\tr{(\calE(X))}$
can be found up to a factor $\pm 1$ using a regularized version of Eq.~[\ref{tr(E(M))}]:
\be\label{tr'}
\tr{ \left( \calE(X) \right) }^2 = C^2 \det{\left( I + M D \right) } \tr{(X)}^2.
\ee
The expression Eq.~[\ref{E(M)}]  does not make sense if $(I+MD)$ is a singular
operator, see a comment~\cite{comment3}.
If this is the case, the Gaussian operator $\calE(X)$ admits only a 
singular representation, as in Eq.~[\ref{Gopr_sing}]. 

Suppose that $\tr{(\calE(X))}=\tr{(X)}$ for any operator $X$, that is
the map $\calE$ is trace preserving (TP). Then it follows from Eq.~[\ref{tr'}]
that a function $f(M)=\det{(I+ MD)}$ does not depend on $M$, as 
long as $M$ is an antisymmetric matrix. From the equality 
$f(D^\dag)=f(-D^\dag)$ we get $\det{(I+D^\dag D)}=\det{(I-D^\dag D)}$ which is
possible only for $D=0$. 
Then it follows from Eq.~[\ref{tr(E(M))}] that
$\calE$ is TP iff
$D=0$ and $C=1$. A transformation of correlation matrices for TP maps
looks particularly simple:
\[
\calE(M)=B M B^T +A.
\]

A TP map $\calE$ preserving an identity operator, $\calE(\Iop)=\Iop$,
is called {\it bistochastic}. Since the identity operator is a Gaussian one
with a zero correlation matrix, we conclude that a Gaussian map $\calE$
is bistochastic iff $A=D=0$, $C=1$. 
A transformation of correlation matrices for bistochastic maps is just
\[
\calE(M)=BMB^T.
\]

\section{Completely positive Gaussian maps}
\label{sec:CP}

Our next goal is to work out conditions under which a Gaussian map
is completely positive  and thus may describe a physical 
transformation of states (may be a conditional transformation such as a 
measurement with a postselection).
By definition, a map $\calE\, : \, \Clif_{2n}\to \Clif_{2n}$ is completely
positive (CP) iff
a map 
\[
\calE\otimes Id \, : \, \Clif_{2n}\otimes \Clif_{2n}\to \Clif_{2n}\otimes
\Clif_{2n}
\]
is positive. Here $Id$ stands for the identity map.
\begin{theorem}\label{theorem:GCP-maps}
A Gaussian map $\calE$ specified  by Eq.~[\ref{Gauss_int},\ref{Action}] is
completely positive iff (i) $C\ge 0$ and (ii) a matrix
\[
M\equiv \left( \ba{cc} A & B \\ -B^T & D \\ \ea \right) 
\]
is real and satisfies  $M^T M \le I$.
\end{theorem}
Before proving the theorem
we shall introduce a fermionic tensor product of linear maps $\otimes_f$
and  a notion of a dual state.
Let $Z\in \Clif_{4n}$ be a monomial in the generators $\cop$. One can always
represent it as $Z=Z'Z''$, where $Z'$ is a monomial in $\cop_1,\ldots,\cop_{2n}$ only and
$Z''$ is a monomial in $\cop_{2n+1},\ldots,\cop_{4n}$ only.
\begin{dfn}\label{def:Z'Z''}
Let  $\calE_1, \calE_2\, : \, \Clif_{2n}\to \Clif_{2n}$ be linear maps.
Define a map $\calE_1\otimes_f\! \calE_2\, : \, \Clif_{4n}\to \Clif_{4n}$ 
on monomials $Z=Z'Z''$ as
\[
(\calE_1\otimes_f\! \calE_2)(Z)=\calE_1(Z') \calE_2(Z''),
\]
and extend it using linearity to the whole algebra $\Clif_{4n}$.
\end{dfn}
In other words, the map $\calE_1\otimes_f\! \calE_2$ acts on the
first half of the modes as $\calE_1$ and on the second
half of the modes as $\calE_2$.
\begin{lemma}\label{lemma:pos}
Let $\calE$ be a parity preserving linear map. Then
$\calE\otimes Id$ is positive iff $\calE\otimes_f\! Id$
is positive.
\end{lemma}
\begin{proof}[Proof:]
Define a Jordan-Wigner transformation
\[
J\, : \, \Clif_{4n}\to \Clif_{2n}\otimes \Clif_{2n}
\]
according to 
\begin{eqnarray}\label{JW}
J(\cop_k)&=&\cop_k\otimes \Pop \quad \mbox{for} \quad 1\le k\le 2n,\\ \nn
J(\cop_k)&=&I\otimes \cop_k \quad \mbox{for} \quad 2n<k\le 4n.\nn
\end{eqnarray}
Here $\Pop$ is the parity operator:
\[
\Pop=i^n \cop_1\cop_2\cdots \cop_{2n}, \quad \Pop^2=I, \quad \Pop^\dag=\Pop.
\]
The definition Eq.~[\ref{JW}] extends to arbitrary operators
by linearity and multiplicativity, $J(XY)=J(X)J(Y)$,
since $J$ preserves commutation relations between the generators.
Thus $J$ is an isomorphism of 
algebras $\Clif_{4n}$ and $\Clif_{2n}\otimes \Clif_{2n}$. Since $J$
preserves  Hermitian conjugation rules, $J(X^\dag)=J(X)^\dag$,
it can be represented as $J(X)=WXW^\dag$, where $W$
is a unitary operator on the space of states of $2n$ fermionic modes
(which is isomorphic to the space of states of two copies of $n$ fermions).
For any monomial $Z=Z'Z''\in \Clif_{4n}$ as above one has
\[
J(Z'Z'')=Z'\otimes \Pop^\ep Z'',
\]
where $\ep=0$ for even $Z'$ and $\ep=1$ for odd $Z'$. Thus
\[
\left((\calE\otimes Id)\circ J \right)(Z)=\calE(Z')\otimes \Pop^\ep Z''.
\]
Besides, according to Definition~\ref{def:Z'Z''} we have
 $(\calE\otimes_f\! Id)(Z)=\calE(Z')Z''$.
Since $\calE$ is parity preserving, $\calE(Z')$ is a linear combination of
monomials having parity $\ep$. Using linearity of $J$ we get
\[
\left(J\circ (\calE\otimes_f\! Id)\right)(Z)=\calE(Z')\otimes \Pop^\ep Z''.
\]
We conclude that
\[
(\calE\otimes Id) \circ J = J\circ (\calE\otimes_f\! Id).
\]
Since $J$ is a unitary isomorphism of algebras, it follows that
positivity of $\calE\otimes Id$ is equivalent to positivity
of $\calE\otimes_f \! Id$.
\end{proof}

To check positivity of the map $\calE\otimes_f\! Id$ we shall make use
of  Jamiolkowski duality between maps and states, see~\cite{Jamiol72,VV02},
slightly adapted to fermions.
\begin{dfn}\label{def:dual}
Let $\calE\, : \, \Clif_{2n} \to \Clif_{2n}$ be a parity preserving
linear map. An operator $\rho_{\calE}\in \Clif_{4n}$ dual to the map $\calE$
is defined as
\[
\rho_{\calE}=(\calE\otimes_f\! Id)(\rho_I),
\]
where
\[
\rho_I=\frac1{2^{2n}}
\prod_{a=1}^{2n} \left(\Iop + i\cop_a \cop_{2n+a}\right)\in \Clif_{4n}.
\]
\end{dfn}
Note that $\rho_I$ is a pure Gaussian state. It can be regarded as 
a 'maximally entangled' state between the modes $\cop_1,\ldots,\cop_{2n}$
and the modes $\cop_{2n+1},\ldots,\cop_{4n}$.
A dual operator completely specifies a linear map. To check that let us
use the isomorphism $\Clif_{4n}\cong \Gr_{4n}$, see
Section~\ref{sec:isom}. Assign Grassmann variables $\theta_1,\ldots,\theta_{2n}$
to the generators
$\cop_1,\ldots,\cop_{2n}$ and variables $\eta_1,\ldots,\eta_{2n}$ to the generators
$\cop_{2n+1},\ldots,\cop_{4n}$. Then one has
\[
\rho_I(\theta,\eta)=\frac1{2^{2n}}\, \exp{\left( i\theta^T\eta\right)}.
\]
It follows that $\rho_{\calE}(\theta,\eta)$ is just a generating function for
the map $\calE$, so by taking derivatives over $\eta$ one can find an image
under $\calE$ of any monomial in $\cop$ generators.

A Gaussian map  defined by Eqs.~[\ref{Gauss_int},\ref{Action}] has a dual
operator 
\[
\rho_{\calE}(\theta,\eta)= \frac{C}{2^{2n}}
\int
\exp{\left[
S(\theta,\zeta) + i\zeta^T\mu + i\mu^T\eta\right]}
\D{\mu}\D{\zeta}.
\]
The integral over $\mu$ gives the integral kernel for the identity map,
see Eq.~[\ref{delta-kernel}]. The integration over $\zeta$ yields
\be\label{g-function}
\rho_{\calE}(\theta,\eta)= \frac{C}{2^{2n}} 
\exp{\left[
S(\theta,\eta)
\right]}.
\ee
Thus Gaussian maps have  Gaussian dual operators.
Now we are ready to prove the main theorem.
\begin{proof}[Proof of Theorem~\ref{theorem:GCP-maps}:]
(a) {\it Necessity:}
If $\calE$ is a CP map, Lemma~\ref{lemma:pos} implies that
$\rho_{\calE}$ must be a non-negative operator. Since $\tr{\rho_{\calE}}=C$, we 
conclude that $C>0$. Since $\rho_{\calE}$ is a self-adjoint operator,
all matrix elements of $M$ must be real. Thus
$\rho_{\calE}$ is proportional to a density operator
of some Gaussian state, see Definition~\ref{def:GS}, that is
$M$ coincides with a correlation matrix of some Gaussian state.
We have already shown in Section~\ref{sec:isom} that its non-negativity
is equivalent to a constraint $M^T M \le I$ on a correlation matrix.

\noindent
(b) {\it Sufficiency:} 
The conditions of the theorem are equivalent
to non-negativity of $\rho_E$. 
To prove complete positivity of $\calE$ 
let us show that it admits a Kraus
representation, that is $\calE(X)=\sum_\alpha A_\alpha X A_{\alpha}^\dag$
for some operators $A_{\alpha}\in \Clif_{2n}$. The proof is a trivial adaptation of
standard ``bosonic'' arguments (see~\cite{VV02} and references therein)
to the case of fermions.

A pure state $|\Phi\ra$  such that $\rho_I=|\Phi\ra\la \Phi|$
(unique up to a phase) satisfies 
\[
\cop_{2n+a}|\Phi\ra=-i\cop_a |\Phi\ra, \quad a=1,\ldots,2n.
\]
It means that an arbitrary pure state $|\Psi\ra$ of $2n$ fermionic modes
can be written as $|\Psi\ra=A|\Phi\ra$ for some operator $A\in \Clif_{4n}$
which involves only the modes $\cop_1,\ldots,\cop_{2n}$ from the first half.
Consider an arbitrary pure state decomposition of $\rho_{\calE}$:
\[
\rho_{\calE}=\sum_\alpha |\Psi_\alpha\ra\la \Psi_\alpha|.
\]
Representing $|\Psi_\alpha\ra= A_\alpha |\Phi\ra$ we get
\[
\rho_{\calE}=\sum_\alpha A_\alpha \rho_I A_{\alpha}^\dag.
\]
Here all $A_{\alpha}$ include only the modes $\cop_1,\ldots,\cop_{2n}$.
Since a dual state completely specifies a map, we conclude that
$\calE(X)=\sum_\alpha A_\alpha X A_\alpha^\dag$ for any operator 
$X\in \Clif_{2n}$.
Complete positivity of $\calE$ follows from 
existence of a Kraus  representation, see~\cite{Choi75}.
\end{proof}

As a simple application of the theorem, let us characterize bistochastic
CP maps. As we already know, $A=D=0$ and $C=1$ for any bistochastic map,
see the last paragraph of Section~\ref{sec:maps}. Then the complete positivity
is equivalent to the matrix $B$ being real and an inequality $B^T B \le I$.
Thus the matrix $B$ can be represented as $B=R_l \tilde{B} R_r$,
where $R_l,R_r\in SO(2n)$ are rotations and $\tilde{B}$ is a diagonal
matrix,
\[
\tilde{B}=\mbox{diag}[B_1,B_2,\ldots,B_{2n}], 
\]
such that $|B_a|\le 1$ for all $a$. The rotations $R_l$ and $R_r$ can be
undone by composing the map $\calE$ with appropriate unitary canonical 
transformations on the left and on the right, see Eq.~[\ref{rotation}].
We conclude that 
\[
\calE=\calE_l\circ \tilde{\calE}\circ \calE_r,
\]
where $\calE_l$, $\calE_r$ are unitary maps and $\tilde{\calE}$ has
an integral representation
\[
\tilde{\calE}(X)(\theta) =  \int \exp{\left[ i\sum_{a=1}^{2n}
 B_a \theta_a \eta_a
 +  \eta_a \mu_a  \right]}
X(\mu) \D{\eta}\D{\mu}.
\]
Evaluation of this integral yields the following action of $\calE$
on the monomials in $\cop$ generators:
\[
\tilde{\calE}(\cop_{a_1}\cdots\cop_{a_k})=B_{a_1}\cdots B_{a_k}  \cop_{a_1}
\cdots\cop_{a_k},
\]
where $k=1,\ldots,2n$.
In other words, $\tilde{\calE}$ is a composition of 
elementary maps 
\[
\cop_a\to B_a\cop_a, \quad |B_a|\le 1, \quad a=1,\ldots,2n,
\]
which all pairwise commute.
It is a direct analogue of a {\it product map} for qubits or bosonic modes.

One can ask whether the set of Gaussian maps is closed under a {\it composition}
of linear maps. As we shall see now, in many important cases
the answer is affirmative. 
\begin{lemma}\label{lemma:compos1}
Suppose $\calE_1$ and $\calE$ are trace preserving Gaussian maps.
Then $\calE_2\circ \calE_1$ is a trace preserving Gaussian map.
\end{lemma}
\begin{proof}[Proof:]
Denote $\calE=\calE_2\circ\calE_1$. It follows from Lemma~\ref{lemma:parity}
that $\calE$ is a parity preserving map. Thus one can define a state
$\rho_{\calE}$ dual to the map $\calE$. Lemma~\ref{lemma:Gmap}
implies that $\rho_{\calE}$ is a Gaussian operator.
Since $\tr{\rho_{\calE}}=1$ we conclude that $\rho_{\calE}$
admits a regular exponential representation as in Eq.~[\ref{g-function}]
and thus a map $\calE$ admits an integral representation Eq.~[\ref{Gauss_int}],
that is $\calE$ is a Gaussian map.
\end{proof}
\begin{lemma}\label{lemma:compos2}
Suppose $\calE_1$ and $\calE_2$ are completely positive Gaussian maps.
Then $\calE_2\circ \calE_1$ is a completely positive Gaussian map.
\end{lemma}
\begin{proof}[Proof:]
The first part of the proof copies the one of the previous lemma.
The dual state $\rho_{\calE}$ is a non-negative operator and thus
$\tr{\rho_{\calE}}>0$ unless $\rho_{\calE}=0$. In the latter case
$\calE\equiv 0$, since a dual state completely specifies a linear map.
The map sending all states to zero is obviously a Gaussian one.  
In the former case one can apply the same proof as above.
\end{proof}

\section{FLO operations as Gaussian maps}
\label{sec:meas}

In this section we put forward an integral representation for a single-mode
occupation number measurement. It will allow to find explicitly
how do the FLO measurements change a correlation matrix of the 
measured state (for each of the two outcomes).

The measurement of the $j$-th mode is described by orthogonal projectors
$\aop_j^\dag\aop_j$ and $\aop_j \aop_j^\dag$ which correspond to the outcomes
``the mode is occupied'' and ``the mode is empty'' respectively.
Let us introduce linear maps $\calE_{j,0}, \calE_{j,1}\, : \,
\Clif_{2n}\to \Clif_{2n}$ describing the adjoint action of the two projectors:
\begin{eqnarray}
\calE_{j,0}(X)&=& (\aop_j \aop_j^\dag)\, X \, (\aop_j \aop_j^\dag),\\ \nn
\calE_{j,1}(X)&=& (\aop_j^\dag \aop_j)\, X \, (\aop_j^\dag \aop_j).
\end{eqnarray}
The index $j$ runs from $1$ to $n$, where $n$ is the total number of fermionic 
modes.  Our first goal is to find operators dual to these maps,
see Definition~\ref{def:dual}:
\[
\rho_{j,0}=(\calE_{j,0}\otimes_f\! Id)(\rho_I) =
(\aop_j \aop_j^\dag)\, \rho_I \, (\aop_j \aop_j^\dag),
\]
and
\[
\rho_{j,1}=(\calE_{j,1}\otimes_f\! Id)(\rho_I) =
(\aop_j^\dag \aop_j)\, \rho_I \, (\aop_j^\dag \aop_j).
\]
Here the operators $\aop_j, \aop_j^\dag$, $j=1,\ldots, n$ are considered as
elements of the Clifford algebra $\Clif_{4n}$, since $\rho_I\in \Clif_{4n}$.
After a simple algebra one gets:
\begin{eqnarray}
\rho_{j,0} &=& \frac1{2^{2n+1}} 
\left( \Iop + i\cop_{2j-1}\cop_{2j} \right)
\left( \Iop - i\cop_{2n+2j-1}\cop_{2n+2j} \right)\nn \\
&& {} \cdot  {\prod_{a}}'
\left( \Iop + i\cop_a\cop_{2n+a} \right). \nn 
\end{eqnarray}
Here the product $\prod'$ is taken over all $a$ in the range $1,\ldots,2n$
excluding $a=2j-1$ and $a=2j$. Analogously one gets
\begin{eqnarray}
\rho_{j,1} &=& \frac1{2^{2n+1}} 
\left( \Iop - i\cop_{2j-1}\cop_{2j} \right)
\left( \Iop + i\cop_{2n+2j-1}\cop_{2n+2j} \right)\nn \\
&& {} \cdot  {\prod_{a}}'
\left( \Iop + i\cop_a\cop_{2n+a} \right). \nn 
\end{eqnarray}
Let us assign Grassmann variable $\theta_1,\ldots,\theta_{2n}$ to
the generators $\cop_1,\ldots,\cop_{2n}$ and variables $\eta_1,\ldots,\eta_{2n}$
to the generators $\cop_{2n+1},\ldots,\cop_{4n}$.
Then the states $\rho_{j,\epsilon}$, $\epsilon \in \{0,1\}$ 
have the following Grassmann representation:
\[
\rho_{j,\epsilon}(\theta,\eta)=\frac1{2^{2n+1}} 
\exp{\left[ S_{j,\epsilon}(\theta,\eta)\right]},
\]
where 
\begin{eqnarray}
S_{j,\epsilon}(\theta,\eta) &=&
i(-1)^{\epsilon} (\theta_{2j-1}\theta_{2j} -  \eta_{2j-1}\eta_{2j}) \nn \\ 
&& {} +i{\sum_{a}}' \theta_a \eta_{a}.\nn
\end{eqnarray}
Here the sum $\sum'$ is taken over all $a$ in the range $1,\ldots,2n$
excluding $a=2j-1$ and $a=2j$. Since a dual state completely specifies a linear
map, it follows from Eq.~[\ref{g-function}] 
that $\calE_{j,\epsilon}$ is a Gaussian map with an integral representation
\be\label{E_jn}
\calE_{j,\epsilon}(X)(\theta)
= \frac12 \int \exp{\left[ S_{j,\epsilon}(\theta,\eta) + i\eta^T \mu \right]}
X(\mu) \D{\eta}\D{\mu}.
\ee
If $X$ is a Gaussian state with a given correlation matrix then $\calE_{j,\epsilon}(X)$
is a Gaussian state whose correlation matrix is given by Eq.~[\ref{E(M)}]. 

Consider as an example a measurement of the first mode and the
outcome ``the mode is empty''. Let $\rho\in \Clif_{2n}$ be the input
Gaussian state  with a correlation matrix $M$,
\[
\rho(\theta) = \frac1{2^n} \exp{\left( \frac{i}2\,
\theta^T M \theta \right)}.
\]
Computing Gaussian integrals in Eq.~[\ref{E_jn}] one gets:
\begin{eqnarray}
(\aop_1\aop_1^\dag)\rho(\aop_1 \aop_1^\dag)(\theta)  &=&
\frac{p_0}{2^n} 
\exp{
\left[ i\theta_1\theta_2 +
\frac{i}2 \sum_{a,b=3}^{2n} L_{ab}\theta_a \theta_b
\right]},  \nn \\
&& \nn \\
L &=& (I - M K)^{-1} M. \nn
\end{eqnarray}
Here $K$ is an antisymmetric $2n\times 2n$ matrix with the only non-zero
matrix elements $K_{12}=-K_{21}=1$, that is
\[
K_{pq}=\delta_{p1}\delta_{q2}-\delta_{p2}\delta_{q1}.
\]
The coefficient $p_0$ turns out to be
\[
p_0=\frac1{2} \Pf{M} \Pf{K - M^{-1}},
\]
which agrees with the earlier derived formula Eq.~[\ref{prob}]
for the probability to observe the mode in the empty state.

In general, if a state $\rho$ to be measured is Gaussian with a given correlation
matrix $M$,
one can use the integral representation Eq.~[\ref{E_jn}]
and the equation Eq.~[\ref{E(M)}] to find a correlation matrix 
$\calE_{j,n}(M)$ of the post-measurement state 
provided that the outcome was $n$. The  probabilities of 
the two outcomes  are given by Eqs.~[\ref{tr(E(M))},\ref{tr'}].

Evolution of a correlation
matrix under unitary elements of FLO (canonical transformations) is
given by Eq.~[\ref{ortogonal}]. Thus the technique developed in the
paper allows to simulate FLO on a classical probabilistic computer
and both aspects of the simulation mentioned in the introduction can
be easily addressed. 

\section{Conclusion}

Notions of a Gaussian state and a Gaussian linear map are
generalized to the case of anticommuting (Grassmann) variables.
Conditions under which a Gaussian map is trace preserving and (or) completely 
positive are explicitly described.
This formalism allows to develop the Lagrangian representation for FLO, which
covers both unitary operations and projectors describing single-mode
measurements. Using the Lagrangian  representation we have reduced classical simulation of
FLO to computation of Gaussian integrals over  Grassmann variables.
Explicit formulas describing evolution of a quantum state under FLO
operations have been put forward. 

\section{Acknowledgments}

Discussions with Alexei Kitaev are gratefully acknowledged.
This work was supported by the National Science Foundation under 
grant number EIA-0086038.



\begin{thebibliography}{100}
\bibitem{comment1} Here we mean the ordinary tensor product of algebras,
such that operators acting only on the first subsystem  commute with operators
acting only on the second subsystem.
\bibitem{comment} One can also define
generalized Gaussian states which are represented by exponents of 
arbitrary quadratic polynomials (with linear terms). We avoid 
using such a definition because (i) Since a parity of the total
number of fermions in the system is preserved by all physical
interactions, only even states are physical; (ii) By adding one
extra fermionic mode one can transform an arbitrary quadratic 
polynomial into an even quadratic polynomial, see~\cite{Knill01}.
\bibitem{comment3} Note that a spectrum of an operator $MD$
coincides with a spectrum of $DM$, so a regularization
$(M^{-1}+D)^{-1}=M(I+DM)^{-1}$ leads to the same 
criterion for $\calE(X)$ to admit a regular representation.
\end{thebibliography}
\end{document}